\documentclass{article}

\usepackage{natbib}      
\usepackage{amsmath, amsthm, amsfonts}
\usepackage{graphicx}    
\usepackage{multirow}    
\usepackage{slashbox}    
\usepackage{rotating}    
\usepackage{color}       
\usepackage{bm}
\usepackage[section]{placeins} 

\def\bSig\mathbf{\Sigma}

\title{Confidence intervals for the area under the receiver operating characteristic curve in the presence of ignorable missing data}

\author{Hunyong Cho$^{1}$,
	Gregory J. Matthews$^{2*}$, and 
	Ofer Harel$^{3}$ \\
	\small $^{1}$Department of Biostatistics, University of North Carolina at Chapel Hill,\\
	\small 135 Dauer Drive, Chapel Hill, NC 27599 \\
	\small $^{2}$Department of Mathematics and Statistics, Loyola University Chicago,\\
	\small 1032 W. Sheridan Road, Chicago, IL  60660\\
	\small $^{3}$Department of Statistics, University of Connecticut,\\
	\small 215 Glenbrook Rd. U-4120, Storrs, CT  06269\\
	\small *email: gmatthews1@luc.edu}

\begin{document}
\maketitle

\abstract{Receiver operating characteristic (ROC) curves are widely used as a measure of accuracy of diagnostic tests and can be summarized using the area under the ROC curve (AUC).  Often, it is useful to construct a confidence intervals for the AUC, however, since there are a number of different proposed methods to measure variance of the AUC, there are thus many different resulting methods for constructing these intervals.  In this manuscript, we compare different methods of constructing Wald-type confidence interval in the presence of missing data where the missingness mechanism is ignorable.  We find that constructing confidence intervals using multiple imputation (MI) based on logistic regression (LR) gives the most robust coverage probability and the choice of CI method is less important. However, when missingness rate is less severe (e.g. less than 70\%), we recommend using Newcombe's Wald method for constructing confidence intervals along with multiple imputation using predictive mean matching (PMM).
	}

\begin{keywords}
	 ROC curve; AUC; Mann-Whitney statistic; confidence interval; multiple imputation; predictive mean matching; missing data, logistic regression
\end{keywords}

\section{Introduction}
\label{s:intro}

Diagnostic tests are often used to classify subjects as either diseased or non-diseased.  Assuming that large values of test results are indicative of diseased status, a subject will be classified as diseased or non-diseased if their test results are above or below a certain threshold, respectively.  Then, for a specific threshold value, the performance of the test can be evaluated using measures such as sensitivity and specificity, where sensitivity is the probability of a true positive or true positive rate (TPR) (i.e. given a positive test result, the subject actually has the disease) and specificity is the probability of a true negative  or true negative rate (TNR) (i.e. given a negative test results, the subject is actually non-diseased).  A more comprehensive way to measure the accuracy of a diagnostic test is to use the receiver-operating characteristic (ROC) curve (Pepe, 2003), which graphically compares specificity and sensitivity levels; specifically, the ROC curve is a plot of sensitivity against {1 - specificity} across all possible cutoff values.  When comparing two diagnostic tests to one another, if the ROC curve for one test is located uniformly higher than the other test across all possible thresholds, the test is said to be strictly more accurate and therefore preferred over the other test.  It is, however, not straight forward to compare two ROC curves that intersect at some location other than the end points.

One common way to summarize the ROC curve is to compute the area under the ROC curve (AUC).  The AUC ranges in value from 0.5 (essentially a meaningless test) to 1 (a perfect test) and can be interpreted as the probability that a randomly chosen diseased subject has a higher test value than that of a randomly chosen healthy subject (given higher test value is indicative of the disease) (Hanley and McNeil, 1982). The AUC is, in most cases, calculated empirically by calculating the area under the sample ROC curve, and it is often of interest to present an interval estimate of the AUC rather than simply a point estimate.

There exist a number of methods that have been proposed to compute an interval estimator for the AUC.  Many of these have been derived based on the equivalence relation between the AUC and Wilcoxon's rank-sum test statistic (Hollander et. al., 2014) such as Bamber (1975), Hanley and McNeil (1982), DeLong et. al. (1988), Mee (1990), and Newcombe (2006). Other proposals, such as Reiser and Guttman (1986) and Newcombe (2006), developed confidence interval estimators of the AUC by assuming a parametric distribution for the test scores and deriving the interval in that manner.  Cortes and Mohri (2005) suggested a confidence interval method using a combinatoric approach and Obuchowski and Lieber (2002) introduced an exact method, which can be used when the estimated AUC is close to 1.

These methods can be easily applied when the true disease status and the test values for all subjects in the sample are known. However, in practice, data may be incomplete.  This often happens when the true disease status of subjects is unknown, which commonly occurs when a patient does not go through a so-called "gold standard test" that, in many cases, is invasive and/or expensive.  Missing data can also occur on patients' test scores for many different reasons including they simply did not have the test performed.  When observations with missing values are removed from analysis (i.e. complete case analysis), the estimator is potentially subject to bias, and when bias is caused specifically by missing disease status, this is referred to as verification bias.

Many corrected estimators of AUC have been proposed to adjust biases caused by missing values. Begg and Greenes (1983), Alonzo and Pepe (2005), He et. al. (2009), along with others, came up with AUC estimators correcting verification bias directly (i.e. ``direct approach").  In order to construct confidence intervals based on those methods, bootstrap techniques are used.  Alternatively, confidence interval methods for complete data settings can be used after applying multiple imputation (MI) techniques to the incomplete dataset (i.e ``MI approach") as in Harel and Zhou (2006), which sought to correct the verification bias for measuring sensitivity and specificity, and Harel and Zhou (2007a).  To address the problem of missing biomarker values, Long et. al. (2011a) and Long et. al. (2011b) came up with a robust approach and an approach using nonparametric MI, respectively. All the estimators correcting for bias introduced here are based on the assumption that the data are missing at random (MAR) (Rubin, 1976), and exploit the covariates to predict the missing values.

Although quite a few estimators have been proposed, relatively little work has been done reviewing the performance of the confidence interval methods for the AUC in the presence of missing data.  In this article, we compare the performance of several Wald-type interval estimators for the AUC in datasets where missing data exist. In our example, the incomplete data will be true disease statuses, and it will be assumed that the missingness mechanism is MAR.

\section{Wald-type confidence interval methods for the AUC}
\label{s:CImethods}
Consider a test such that higher test scores are associated with a higher probability that the subject is diseased and vice versa.  By definition, the AUC ($\theta$) is the integral of the TPR of the theoretical ROC curve by its false positive rate (FPR = 1 - TNR), that is, $\theta = \int_{-\infty}^{\infty}TPR(t)dFPR(t),$ where TPR(t) and FPR(t) are the TPR and FPR for a threshold value $t$. As the AUC can be interpreted as the probability that a randomly chosen diseased subject has higher biomarker value than a randomly chosen healthy subject's biomarker value (plus half of the probability of ties is added, if any), it can also be expressed as: $\theta = P[Y > X] + \frac{1}{2}P[Y = X],$ where $X$ and $Y$ are the biomarker of the randomly chosen non-diseased and diseased subjects respectively. So it is straightforward to have an unbiased estimator of the AUC as: $$\hat{\theta} = \frac{\sum_{i=1}^{n_Y} \sum_{j=1}^{n_X} \{I(y_i > x_j) + \frac{1}{2}I(y_i = x_j)\}}{n_{Y}n_{X}},$$ where $x_j$ and $y_i$ are the test scores for the $j$-th and $i$-th individual in the non-diseased and diseased group, respectively, and j = 1, ..., $n_X$ and i = 1, ..., $n_Y$.

As the variance estimation of the AUC is not obvious, various variance estimators of the AUC have been proposed. Given a variance estimator, confidence intervals (CI's) can be built in several ways. The simplest and the most commonly used form is the Wald-type interval. This type of CI relies on large sample theory and evaluates the standard error at a point estimate to have upper and lower bounds. Another popular method for CI construction is the score-type (Wilson and Hilferty, 1929) interval. This method is also an asymptotic approach, but the standard errors are evaluated at the candidates of the confidence bounds, not at the point estimate only. Other types of confidence interval constructions include likelihood-ratio method and exact methods.

Despite some drawbacks with Wald-type intervals, such as inappropriateness to use when the sample is small or when the population parameter for proportional data are extreme, this type of interval is computationally simple and applicable to almost all types of variance estimators. Additional assumptions should be posed to get other types of intervals other than Wald-type intervals under missing data settings. In this article, we consider the Wald-type CI's only, presenting some of them below.

\subsection{Bamber's method}
\label{ss:Bamber}
Bamber (1975) presented the variance estimator of $\hat{\theta}$ using the equivalence relation between the Mann-Whitney statistic ($U$) and the AUC ($\theta$) that $\theta = \frac{U}{n_{X}n_{Y}}$ and the variance of $\hat{U}$ (Noether, 1967):
\begin {equation*}
\begin {split}
\widehat{V(\hat{\theta})} =\frac{1}{4(n_{X}-1)(n_{Y}-1)} \bigg[ & p(X \neq Y) + (n_X - 1)b_{XXY}
+ (n_Y - 1)b_{YYX} \\
&- 4(n_X + n_Y - 1)(\hat{\theta} - \frac{1}{2})^2 \bigg],
\end {split}
\end {equation*} where 
$p(X \neq Y) = \frac{\sum_{i=1}^{n_Y} \sum_{j=1}^{n_X} I(y_i \neq x_j)}{n_{Y}n_{X}}$,
$b_{XXY} = \\ \frac{\sum_{i=1}^{n_Y} [u_{i.}(u_{i.} - 1) + v_{i.}(v_{i.} - 1) - 2u_{i.}v_{i.}]}{n_X(n_X - 1)n_Y}$,
$b_{YYX} =$ $\frac{\sum_{j=1}^{n_X} [u_{.j}(u_{.j} - 1) + v_{.j}(v_{.j} - 1) - 2u_{.j}v_{.j}]}{n_Y(n_Y - 1)n_X}$,
$v_{.j} = \sum_{i=1}^{n_Y} [I(y_i > x_j) + \frac{1}{2}I(y_i = x_j)]$,
$u_{.j} = n_Y - v_{.j}$,
$v_{i.} = \sum_{j=1}^{n_X} [I(y_i > x_j) + \frac{1}{2}I(y_i = x_j)]$,
$u_{i.} = n_X - v_{i.}$ and $X_1, X_2, Y_1$ and $Y_2$ are random variables sampled independently without replacement from X and Y.

Here $b_{XXY}$ and $b_{YYX}$ are an unbiased estimator of
$ B_{XXY} = P(X_1, X_2 < Y) + P(Y < X_1, X_2) - P(X_1 < Y < X_2) - P(X_2 < Y < X_1)$ and
$ B_{YYX} = P(Y_1, Y_2 < X) + P(X < Y_1, Y_2) - P(Y_1 < X < Y_2) - P(Y_2 < X < Y_1)$, respectively.

Using this variance estimator, a $1-\alpha$ CI for the AUC can be constructed as $\hat{\theta} \pm z_\frac{\alpha}{2}\sqrt{\widehat{V(\hat{\theta})}}$, where $z_\frac{\alpha}{2}$ is the upper $\frac{\alpha}{2}$ percentile of the standard normal distribution.

\subsection{Hanley-McNeil's method I}
\label{ss:HM1}
Hanley and McNeil (1982) extended the ideas in Bamber (1975) to simpler variance estimators. Unlike Bamber's variance estimator, those of Hanley and McNeil (and Newcombe as well discussed in later subsection) are not unbiased by using $n_X$ and $n_Y$ in the denominator. As Bamber (1975) did for his method, we will use $n_X - 1$ and $n_Y - 1$ instead in the denominator to make the estimators unbiased for this method and for the later methods (\ref{ss:HM1} - \ref{ss:NW}). The under-coverage of the methods in Hanley and McNeil (1982), which were discussed in Newcombe (2006) is partly due to the underestimation of the variance. 

Another shortcoming of the variance formula in Hanley and McNeil (1982) is that it is appropriate only when the biomarker is sufficiently continuous so that there are no ties between the diseased and the non-diseased. In many cases, however, the biomarkers are discrete, or even when they are continuous, ties often happen by rounding values to significant digits. In our simulation study, we apply a generalized formula that incorporates the tie situations.

Then the revised variance estimator is as follows:
\begin{equation*}
\begin{split}
\widehat{V(\hat{\theta})} = \frac{1}{(n_{X}-1)(n_{Y}-1)} \bigg[ & \hat{\theta}(1-\hat{\theta}) -\frac{1}{4}p(Y=X)
+ (n_Y - 1)(\hat{Q}_1 - \hat{\theta}^2) \\
& + (n_X - 1)(\hat{Q}_2 - \hat{\theta}^2) \bigg],
\end{split}
\end{equation*} 
where $\hat{Q}_1=\frac{\sum_{j=1}^{n_X} [\sum_{i=1}^{n_Y} \{ I(y_i > x_j) + \frac{1}{2}I(y_i = x_j) \} ]^2}{n_{X}n_{Y}^{2}}$ is an estimator of 
$Q_1 = P(Y_1, Y_2 > X) + \frac{1}{2} P(Y_1 > Y_2 = X) + \frac{1}{2} P(Y_2 > Y_1 = X) + \frac{1}{4} P(Y_1 = Y_2 = X)$, 
or the probability that biomarkers of two randomly chosen diseased subjects, possibly the same subjects, are greater than or equal to that of a randomly chosen non-diseased subject with half the weight to the equality. Similarly, $\hat{Q}_2=\frac{\sum_{i=1}^{n_Y} [\sum_{j=1}^{n_X} \{ I(y_i > x_j) + \frac{1}{2}I(y_i = x_j) \} ]^2}{n_{X}^2n_{Y}}$ is an estimator of 
$Q_2 = P(Y > X_1, X_2) + \frac{1}{2} P(Y = X_1 > X_2) + \frac{1}{2} P(Y = X_2 > X_1) + \frac{1}{4} P(Y = X_1 = X_2)$. 
(The mathematical proof that the revised estimator is an unbiased estimator is provided in the Appendix.)

The Wald-type interval is given as $\hat{\theta} \pm z_\frac{\alpha}{2}\sqrt{\widehat{V(\hat{\theta})}}$.

\subsection{Hanley-McNeil's method II}
\label{ss:HM2}
Hanley and McNeil (1982) further simplified the variance estimator by assuming that X and Y are exponentially distributed: 
\begin{equation*} \label{eq:HM1}
\begin{split}
\widehat{V(\hat{\theta})} = \frac{1}{(n_{X}-1)(n_{Y}-1)} \bigg[ & \hat{\theta}(1-\hat{\theta}) -\frac{1}{4}p(Y=X) + (n_Y - 1)(\hat{Q}_1 - \hat{\theta}^2)\\
&+ (n_X - 1)(\hat{Q}_2 - \hat{\theta}^2) \bigg],
\end{split}
\end{equation*}
where $\hat{Q}_1=\frac{\hat{\theta}}{2-\hat{\theta}}$ and $\hat{Q}_2=\frac{2\hat{\theta}^2}{1+\hat{\theta}}$.

\subsection{Newcombe's Wald method}
\label{ss:NW}
Newcombe (2006) suggested a modification to the Hanley-McNeil's method II by replacing both $n_X$ and $n_Y$ in the numerator with $N=\frac{n_X + n_Y}{2}$. Then the variance estimator becomes: $$\widehat{V(\hat{\theta})}=\frac{\hat{\theta}(1-\hat{\theta})}{(n_{X}-1)(n_{Y}-1)}\bigg[ 2N - 1 - \frac{3N-3}{(2-\hat{\theta})(1+\hat{\theta})} \bigg]$$

\subsection{DeLong's method}
\label{ss:DL}
Based on the method in Sen (1960), DeLong et. al. (1988) derived a covariance matrix estimator of $\hat{\theta}$ which makes possible nonparametric comparisons of two or more ROC curves. Still the idea can be applied to construct a Wald-type CI for the area under a single ROC curve: $$\widehat{V(\hat{\theta})} = \frac{\hat{s}_{10}}{n_X} + \frac{\hat{s}_{01}}{n_Y},$$ 
where $\hat{s}_{10} = \frac{1}{n_X - 1} \sum_{j = 1}^{n_X} [\frac{v_{.j}}{n_Y} - \hat{\theta}]^2$ 
and $\hat{s}_{01} = \frac{1}{n_Y - 1} \sum_{i = 1}^{n_Y} [\frac{v_{i.}}{n_X} - \hat{\theta}]^2$.
Using this variance estimator, a Wald-type CI can be constructed as: 
$\hat{\theta} \pm z_\frac{\alpha}{2}\sqrt{\widehat{V(\hat{\theta})}}$.

\section{Missing data and multiple imputation}
\label{s:MImethods}
A common problem in applied statistics is incomplete data. Incomplete data occur in many medical, social, and demographic analyses and leads to difficulties in performing inference. One way of handling the missing data problem is multiple imputation (Rubin, 2004, Harel and Zhou, 2007b). MI is a Monte Carlo technique that involves three steps: imputation, analysis and combining of results. First, the missing values are filled in by $m$ simulated values to create $m$ completed datasets where $m$ is commonly larger than 5. The imputation model consists of selecting a plausible probability model on both observed and missing values to simulate values for the missing observations. The imputation model should be selected to be compatible with the analysis to be performed on the imputed datasets where the analysis model is determined by the statistical problem in question. Following this, each of the $m$ imputed datasets is analyzed separately using some complete data statistical method of interest. This is then followed by the results of each of the $m$ analyses being combined (Rubin, 2004) to produce point and/or interval estimates that incorporate both between and within imputation variability.

\subsection{Missingness mechanism assumptions}
\label{ss:MAR}
Define the missingness mechanism to be $R$ ,which is a random variable that is a 1 if the value is missing and 0 otherwise.  In order to perform MI, assumptions about the missingness mechanism must be made. The simplest missingness assumption is that the data are missing completely at random (MCAR).  This type of missingness implies that the probability of an observation does not depend on the observed or unobserved values in the data. Data are said to be missing at random (MAR) if the probability of missingness does not depend on the unobserved data (Little and Rubin, 2002).  If the probability of missingness is related to unobserved values in the dataset, the missingness is said to be missing not at random (MNAR).  In this case, to perform MI, the missingness mechanism must be explicitly modeled.  In the case of MAR and MCAR, if the parameters of the data model and the parameters of the missing data model are distinct, then missingness mechanism, $R$, does not need to be modeled and the missing data mechanism is referred to as ignorable.  In this manuscript, we focus solely on data whose missingness mechanism is ignorable.

\subsection{Multiple imputation techniques}
\label{ss:MI}
One common, but simple method for performing MI is to model the data as if it comes from a multivariate normal distribution. Once this model is fit, imputed values are drawn from the posterior predictive distribution using data augmentation (DA) which is a kind of MCMC algorithm (Schafer, 1997).  A stand alone statistical package named ``NORM" (NORM, 1999) is available for performing this MI method.  Additionally, there is a package for performing this method (Novo, 2013) in R (R, 2017).

While the previous method is based on multivariate normal distribution, in many applied statistical problems the adequate imputation model is not multivariate normal. For example, a dataset concerning the Alzheimer's disease (AD) that we will analyze in Section \ref{s:example} contains the variable outcome with a value of 1 if the patient has AD or 0 if the patient does not have the disease.  Bernaards et. al. (2007) explored using a multivariate normal model when the data clearly violated that assumption, namely in the presence of a binary variable.  In that manuscript, they imputed under the multivariate normal model and used a variety of rounding techniques to convert the continuos imputed values back to binary variables.  Here, we implement one of these methodologies in our analysis specifically the adaptive rounding approach.  Adaptive rounding is a rounding procedure where the threshold value for rounding to zero or one is based on normal approximation to the binomial distribution. The adaptive rounding threshold is given by $\bar{\omega} - \sqrt{\bar{\omega}(1 - \bar{\omega})}\Phi^{-1}(\bar{\omega})$ where $\bar{\omega}$ is the imputed variable obtained from the multivariate normal imputation procedure, and $\Phi(\cdot)$ is the cumulative density function of the standard normal distribution.  For more information on rounding in multiple imputation see Demirtas (2009).  

Multivariate imputation by chained equations (MICE) (White et. al., 2011, van Buuren, 2012, van Buuren et. al., 2017) is another widely used MI method. In MICE's sequential process, a joint distribution for the imputation models does not need to be explicitly specified, and thus makes this method very flexible (Allison, 2009). Despite the lack of theoretical justification as to whether MICE's sequential model converges to the distributions of the missing variables, it was demonstrated to perform well in many simulation settings (van Buuren et. al., 2006). Zhu and Raghunathan (2016) claim that under certain conditions such as valid model specification or good fit of distribution to data, sequential regression approach may work well even with models that are incompatible with the conditional distributions.

While there are numerous ways to implement MICE, in this paper we focus on predictive mean matching (PMM) (Little, 1988) and logistic regression approach (LR) (Rubin, 2004) since the variable with missingness in our simulation study is dichotomous.  Here, the MICE framework is implemented in R via the ``mice" package (van Buuren and Groothuis-Oudshoorn, 2011).

\subsection{Multiple imputation combining rules}
\label{ss:Combining Rule}
Once MI is performed, each of the $m$ imputed datasets are analyzed to produce estimates ($\hat{Q}_i$) of the quantity of interest ($Q$) and estimates ($\hat{V}_i$) of the associated variance ($V$), where $i=1,2,...,m$. Assuming that the 
sampling distribution of $Q$ is normally distributed then, according to Rubin's combining rules, 
$\frac{Q - \bar{Q}}{\sqrt{W + \frac{m+1}{m} B}} \sim t_\nu,$
where $\bar{Q}=\frac{1}{m}\sum_{i=1}^{m}\hat{Q}_i$, $W=\frac{1}{m}\sum_{i=1}^{m}\hat{V}_i$, $B= \frac{1}{m-1} \sum_{i=1}^{m}(\hat{Q}_i-\bar{Q})^2$, and $\nu=(1+\frac{m}{m+1}\frac{W}{B})^2(m-1)$.

\subsection{Making inference about the AUC with MI}
\label{s:AUCMI}
MI techniques can be employed to fill in the missing values to make an inference on the AUC. From the multiple ($m$) sets of imputed data, two vectors of statistics are obtained: $\bm{\hat{\theta}} = \big\{\hat{\theta}_{(i)}| i = 1, 2, ..., m\big\}$ and $\bm{\hat{V}} = \big\{\widehat{V(\hat{\theta})}_{(i)}| i = 1, 2, ..., m\big\}$, where $\hat{\theta}_{(i)}$ and $\widehat{V(\hat{\theta})}_{(i)}$ are the estimate and the variance estimate of the AUC for the $i^{th}$ imputed data. The variance estimates of the AUC can be obtained by applying one of the methods mentioned in Section~\ref{s:CImethods}.
Given that the sample size is large, the sample AUC's are asymptotically normal. Then these vectors are combined to form a 95\% CI, $\bar{\theta} \pm t_{\nu,.975} \sqrt{\hat{V}_*},$ where $\bar{\theta} = \frac{1}{m}\sum_{i=1}^{m}{\hat{\theta}_{(i)}}$, $\hat{V}_* = W + \frac{m + 1}{m}B$, 
$W = \frac{1}{m} \sum_{i=1}^{m}{\widehat{var(\hat{\theta})}_{(i)}}$, $B = \frac{1}{m-1}\sum_{i=1}^{m}(\hat{\theta}_i-\bar{\theta})^2$, 
and $\nu = [1 + (\frac{m}{m+1})\frac{W}{B}]^2 (m-1)$.

\section{Simulation Study}
\label{s:sim}
Simulation-based methods are quite often used to evaluate performance in incomplete data settings.  For example, Mitra and Reiter (2016), Demirtas and Hedeker (2008), Bernaards et. al. (2007), Demirtas (2007), Demirtas and Hedeker (2007), and Demirtas et. al. (2007) are a few of the numerous examples of such research.  Here we follow these examples and use simulation-based methods to study the performance of different Wald-type confidence intervals for the AUC in the presence of missing data.  Specific details of the simulation follow.   

We generated simulated data with 8 variables: disease status (D), biomarker (T), 5 covariates (Z = $(Z_1, Z_2, Z_3, Z_4, Z_5)^{\prime}$), and missingness indicator (R). Given a set of random covariates (Z), D, T, and R were randomly drawn according to the parameters: the prevalence rate ($\phi$), the AUC ($\theta$), and the rate of missing observations (or the missing coverage, $\rho$) sequentially. The generating scheme for D, T, and R is largely from Alonzo and Pepe (2005). Three different sample sizes, 50, 100, and 200, were considered for each case. For each combination of parameters and sample size, a simulation with 10,000 replicates was performed. This simulation study was done using the software R 3.4.2 (R Core Team, 2017). The code is provided as supplementary material.

After creating random data, $95\%$ CI's were constructed using 3 different MI techniques (PMM, LR, and NORM) and 5 different CI methods. The complete data were also analyzed for comparison with the MI. Then they were evaluated by measuring the coverage probability, the left and right non-coverage probability, and the average confidence interval length.

\subsection{Distributional assumptions}
The covariates are assumed to be multivariate normal:
$Z \sim MVN(\mu_Z, \Sigma_Z),$
where $\mu_Z$ is a vector containing the means of 5 covariates, and $\Sigma_Z$ is a 5 by 5 covariance matrix of $Z$.

The disease status, D, is 1 for diseased subjects and 0 for non-diseased subjects. It was generated by random Bernoulli draw with prevalence rate $\phi$, which is a function of linear combination of $Z$'s:
$D|Z \sim Bern(\phi),$ where $logit(\phi) = \alpha_0 + \alpha_1^{\prime} Z$.

The biomarker, T, is normally distributed with its mean conditional on the values of the disease status and the covariates:
$T|D,Z \sim N(\mu_{T}, \sigma_T^2),$ where $\mu_{T} = \beta_0 + \beta_1 D + \beta_2^{\prime}Z + \beta_3^{\prime}DZ$ where $\beta_2^{\prime}$ and $\beta_3^{\prime}$ are vectors of regression coefficients both containing 5 elements.  

The missingness indicator on disease status, R, is 1 for non-observed subjects and 0 for observed subjects. If a subject has either a biomarker value within upper $q_1$-th quantile, or at least one of the covariates is within its $q_2$-th quantile, then he/she always goes through the golden standard test to verify his/her true disease status. Otherwise, the probability of non-verification was set as $0 < \gamma < 1$; $P(R = 0 |T > t^{q_1}$  or  $Z_i > z_i^{q_2}$ for some $1 \le i \le 5) = 1$, $P(R = 1 |T \le t^{q_1}$  and  $Z_i \le z_i^{q_2}$ for all $1 \le i \le 5) = \gamma$, where $t^{q_1}$ denotes the $q_1$-th quantile of the distribution of T and $z_i^{q_2}$ denotes the $q_2$-th quantile of the distribution of $Z_i$.

\subsection{Parameter specification}
We set the mean of $Z$ as $\mu_Z = 0_5$, where $k_5$ denotes a column vector $(k, k, k, k, k)^{\prime}$ for some number $k$, and the covariances of $Z$ as 
$$
\Sigma_Z = 
\Bigg( \begin{smallmatrix}
1 & 0 & 0.3 & 0.4 & -0.4 \\
0 & 1 & 0.2 & 0.2 & 0 \\
0.3 & 0.2 & 1 & 0.7 & -0.5 \\
0.4 & 0.2 & 0.7 & 1 & -0.2 \\
-0.4 & 0 & -0.5 & -0.2 & 1
\end{smallmatrix}\Bigg).$$
$\alpha_0$ is set as 0 to have $E(\phi)$ = 0.5, and as 1.6111 to have $\phi \in 0.70000 \pm 9.0 \times 10^{-6}$ with 95$\%$ confidence. We also set $\alpha_1 = 1_5$,
$\beta_0 = 0$, $\beta_2 = 0.1_5$, $\beta_3 = 0.05_5$, and $\sigma_T^2 = 1$. 

By setting $\sigma_T^2$ as a constant, rather than a function of $D$, the biomarker values are assumed to be almost homoscedastic between the diseased and the non-diseased groups. The reason why homoscedasticity is addressed is because heteroscedasticity inflates the variance of the sampling distribution of the AUC, and thus lowers the coverage probabilities of the CI's. In our simulation study, the variability of the biomarker, $T$, is slightly higher for the disease group than that for the non-diseased group. However, the disparity in variance between the groups is minor and therefore we did not consider the two groups of biomarkers to be heteroscedastic.

Given those parameters and $\alpha_0=0$, $\beta_1$ = 0.8089, 1.4486, 1.9767 and 2.9670 give the desired values of AUC ($\theta$) = 0.8, 0.9, 0.95 and 0.99. For $\alpha_0=1.6111$, the values of $\beta_1$ = 0.8319, 1.4729, 2.0019 and 2.9939 make $\theta$ to be 0.8, 0.9, 0.95 and 0.99.

If we set $\gamma = 0.9$, $q_1 = 0.85$, and $q_2 = 0.9$, the missing percentage ($\rho$) is roughly $50\%$. If we set $\gamma = 0.95$, $q_1 = 0.9$, and $q_2 = 0.9$, the missing coverage ($\rho$) is roughly $70\%$.  If we set $\gamma$ = 0.95, q1 = 0.99, and q2 = 0.99, the missing coverage ($\rho$) is roughly 90\%. 

To analyze complete data, the missingness indicator (R) was ignored. To get CI's for incomplete data, MI techniques were applied to the same data with missingess indicator (R) considered.  Imputations were performed m = 10 times with 5 iterations each for all cases. For NORM imputation, a non-informative prior for the multivariate normal distribution was assumed and the iterations started from estimates by the expectation-maximization (EM) algorithm.

For each of 72 different settings (i.e. four AUC values, two prevalence rates, three missingness rates, and three sample sizes), 10,000 simulation replicates were performed for each of three imputation methods (PMM (MICE), LR (MICE), and NORM).

\subsection{Evaluation}
We evaluate the performance of CI's by measuring coverage probability (CP), left and right non-coverage probability (LNCP and RNCP), and the confidence interval length (CIL). Coverage probability is defined as the proportion of CI's that capture the population AUC and the proportion of CI's of which upper (or lower) limit lies below (or above) the population AUC is the LNCP (or RNCP). 
For calculating the CIL the confidence intervals were truncated above at 1 and below at 0.  Each evaluation measure is presented by being averaged across some parameters and/or across simulation replicates if necessary.

For CP, the smaller the error (i.e. difference between the actual and nominal CP) is, the better performing the CI's are. Given the same confidence level, shorter CIL's are preferred. LNCP and RNCP are considered good if they are balanced. To additionally see the stability of the coverage statistic, that is, to see if there is a chance that seemingly good average CP happens to result from averaging extremely small and large CP's, we also measured the mean absolute error of the CP (MAE): $MAE = \sum_{i=1}^{n_s} \frac{| CP_i - CP |} {n_s}$, where CP is the nominal coverage probability, $CP_i$ is the actual coverage probability of the $i^{th}$ setting, and $n_s$ is the number of settings.

\section{Results}
\label{s:results}
For each imputation technique, CI method and missing coverage ($\rho$), the average CP, the MAE of CP and the average CIL were calculated. Table~\ref{tab:performance2} shows the results when $\rho = 70\%$. (Results for values of $\rho=50\%$ and $\rho=90\%$ can be seen in Tables~\ref{tab:performance1} and \ref{tab:performance3}, respectively, in the appendix). Table~\ref{tab:NCP2} presents the results about the average LNCP and RNCP for $\rho = 70\%$.  (LNCP and RNCP results for $\rho=50\%$ and $\rho=90\%$ can be found in Tables~\ref{tab:NCP1} and \ref{tab:NCP3}, respectively, in the appendix.)  For each table, the first five rows correspond to the full dataset that has no missing values, the second five rows are for complete case analysis (i.e. naive analysis), and the remaining rows are for incomplete datasets after applying multiple imputation by PMM, LR, and NORM, respectively.

\begin{figure}[h]
	\centering
	\includegraphics[width=\textwidth]{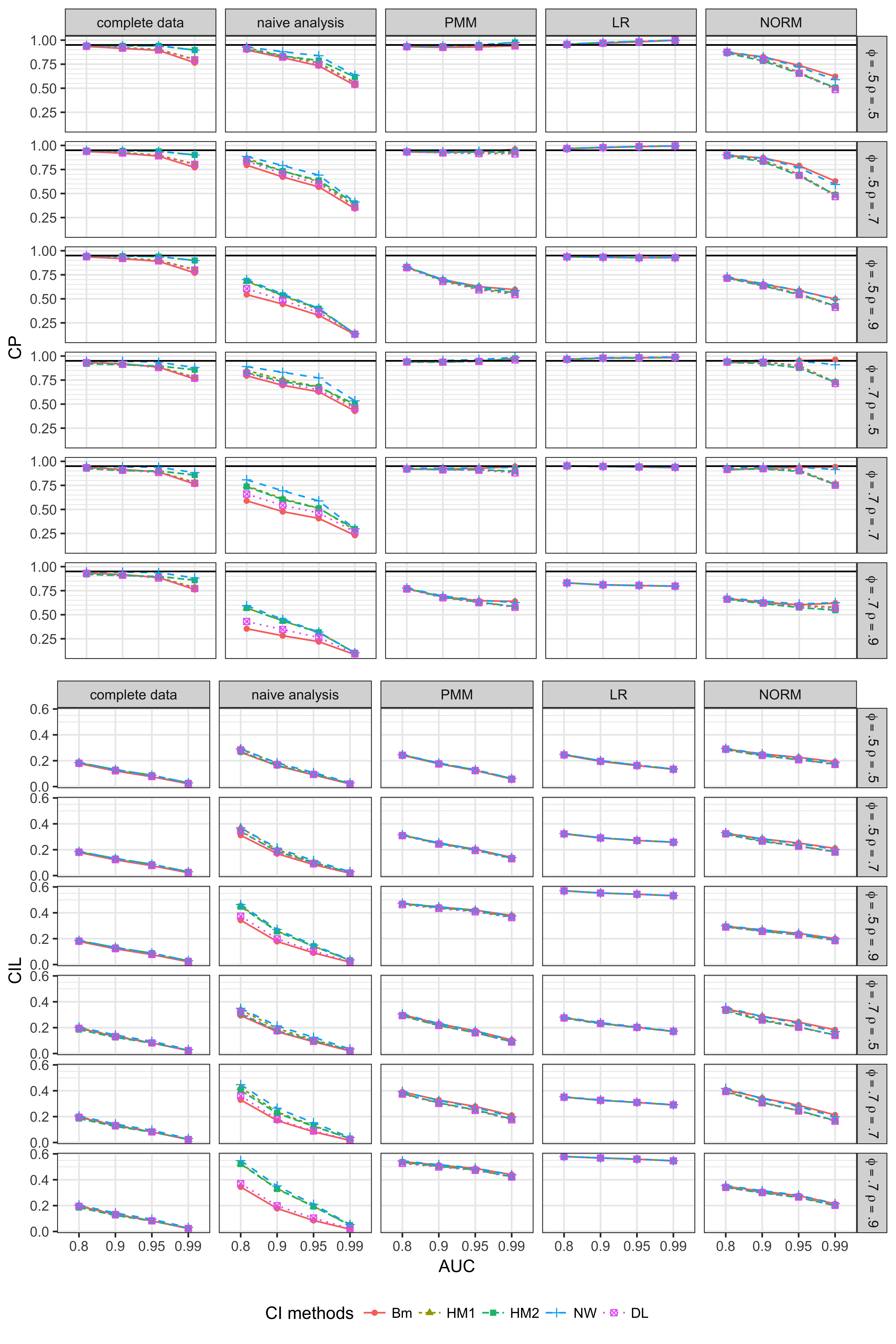}
	\caption{CP and CIL by MI techniques, CI methods, $\phi$ and $\rho$}   \label{fig:phirho}
\end{figure}

\begin{figure}[h]
	\centering
	\includegraphics[width=\textwidth]{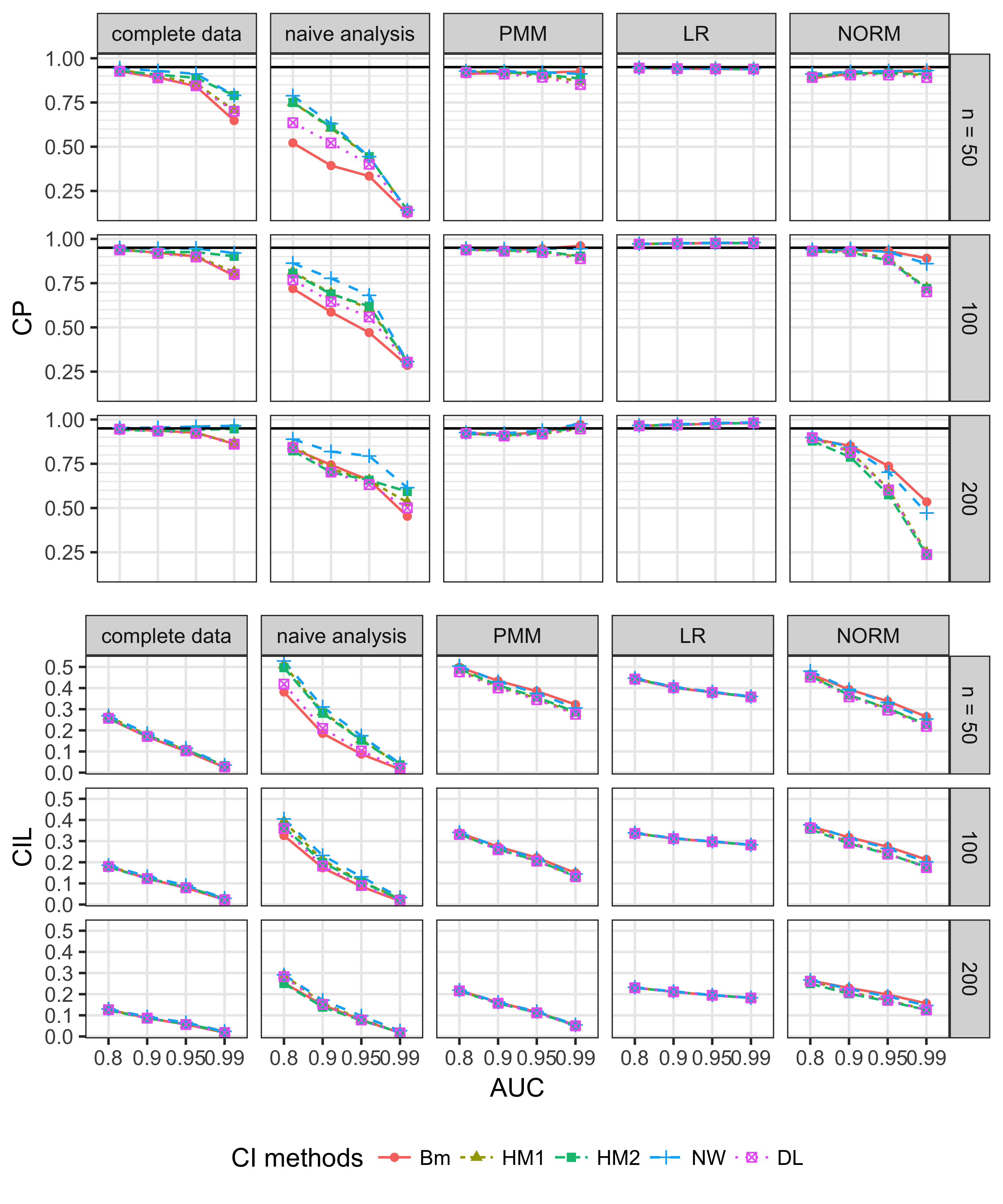}
	\caption{CP and CIL by MI methods, CI methods and $n$ when $\rho = 70\%$}  \label{fig:CPCIL}
\end{figure}

The average CP and the average CIL are also shown in Figure~\ref{fig:phirho} by AUC levels ($\theta$), MI techniques (compared with analysis on data that have no missing value), the prevalence rate ($\phi$) and the missing coverage ($\rho$). Then the performances across different values of sample size ($n$) when $\rho = 70\%$ are presented in Figure~\ref{fig:CPCIL}.

We first look at the results for the complete datasets, and see how the performance changes when we do complete case analysis for incomplete data. Then we move on to different imputation techniques for the incomplete data with moderate level of missingness ($\rho = 70\%$).

\begin{table}[ht]
	\caption{Performance of CI's for each imputation and CI method when $\rho = 70\%$} \label{tab:performance2}
	\centering
	\begin{tabular}{llcccc cccc cccc}
		\hline
		MI & CI &
		\multicolumn{4}{c}{CP} &
		\multicolumn{4}{c}{MAE (CP)} &
		\multicolumn{4}{c}{CIL} \\
		\multicolumn{2}{r}{}  & $\theta$ = 0.8 & 0.9 & 0.95 & 0.99  & 0.8 & 0.9 & 0.95 & 0.99 & 0.8 & 0.9 & 0.95 & 0.99 \\ 
		\hline
		\multirow{5}{*} {\begin{turn}{90} complete\end{turn}} & Bm & .938 & .917 & .890 & .767 & .012 & .033 & .060 & .183 & .187 & .126 & .079 & .022 \\ 
		& HM1 & .942 & .920 & .895 & .795 & .008 & .030 & .055 & .155 & .192 & .130 & .082 & .024 \\ 
		& HM2 & .936 & .923 & .919 & .880 & .015 & .027 & .034 & .075 & .186 & .128 & .084 & .028 \\ 
		& NW & .949 & .943 & .940 & .892 & .004 & .010 & .017 & .068 & .195 & .137 & .091 & .030 \\ 
		& DL & .937 & .914 & .888 & .787 & .013 & .036 & .062 & .163 & .188 & .127 & .080 & .023 \\ 
		\hline
		\multirow{5}{*} {\begin{turn}{90} naive\end{turn}} & Bm & .692 & .575 & .488 & .287 & .258 & .375 & .462 & .663 & .320 & .169 & .084 & .017 \\ 
		& HM1 & .806 & .673 & .568 & .327 & .144 & .277 & .382 & .623 & .398 & .218 & .116 & .027 \\ 
		& HM2 & .793 & .668 & .574 & .347 & .157 & .282 & .376 & .603 & .371 & .206 & .112 & .027 \\ 
		& NW & .847 & .742 & .640 & .355 & .103 & .208 & .310 & .595 & .408 & .238 & .134 & .034 \\ 
		& DL & .749 & .623 & .531 & .313 & .201 & .327 & .419 & .637 & .354 & .181 & .090 & .019 \\ 
		\hline
		\multirow{5}{*} {\begin{turn}{90} PMM \end{turn}} & Bm & .925 & .923 & .931 & .953 & .025 & .027 & .020 & .021 & .352 & .290 & .241 & .175 \\ 
		& HM1 & .929 & .920 & .916 & .905 & .022 & .030 & .034 & .051 & .345 & .275 & .223 & .155 \\ 
		& HM2 & .929 & .922 & .922 & .914 & .023 & .028 & .028 & .047 & .346 & .277 & .226 & .158 \\ 
		& NW & .933 & .930 & .935 & .944 & .020 & .022 & .015 & .023 & .355 & .289 & .237 & .168 \\ 
		& DL & .926 & .916 & .912 & .896 & .024 & .034 & .038 & .060 & .341 & .272 & .221 & .153 \\ 
		\hline
		\multirow{5}{*} {\begin{turn}{90} LR \end{turn}} & Bm & .960 & .962 & .964 & .964 & .023 & .028 & .034 & .035 & .337 & .308 & .290 & .274 \\ 
		& HM1 & .961 & .963 & .965 & .967 & .023 & .029 & .034 & .036 & .338 & .309 & .291 & .275 \\ 
		& HM2 & .961 & .963 & .965 & .966 & .023 & .029 & .034 & .036 & .337 & .309 & .291 & .275 \\ 
		& NW & .962 & .964 & .967 & .968 & .024 & .030 & .035 & .037 & .340 & .311 & .293 & .275 \\ 
		& DL & .960 & .962 & .965 & .966 & .023 & .028 & .034 & .036 & .336 & .308 & .290 & .274 \\ 
		\hline
		\multirow{5}{*} {\begin{turn}{90} NORM \end{turn}} & Bm & .902 & .901 & .863 & .785 & .048 & .053 & .095 & .179 & .367 & .314 & .270 & .211 \\ 
		& HM1 & .910 & .887 & .804 & .624 & .040 & .066 & .151 & .326 & .363 & .290 & .238 & .176 \\ 
		& HM2 & .903 & .875 & .790 & .622 & .047 & .077 & .164 & .328 & .355 & .286 & .237 & .177 \\ 
		& NW & .916 & .904 & .852 & .754 & .036 & .055 & .108 & .208 & .375 & .311 & .263 & .201 \\ 
		& DL & .907 & .883 & .797 & .610 & .043 & .069 & .156 & .340 & .359 & .286 & .235 & .173 \\ 
		\hline
	\end{tabular}
\end{table}

  \begin{table}[ht]
  	\caption{Non-coverage probabilities of CI's for each imputation and CI method when $\rho = 70\%$} \label{tab:NCP2}
  	\centering
  	\begin{tabular}{llcccc p{0.01cm} cccc}
  		\hline
  		MI & CI &
  		\multicolumn{4}{c}{LNCP} &&
  		\multicolumn{4}{c}{RNCP} \\
  		\multicolumn{2}{r}{}  & $\theta$ = 0.8 & 0.9 & 0.95 & 0.99  && 0.8 & 0.9 & 0.95 & 0.99 \\ 
  		\hline
  		\multirow{5}{*} {\begin{turn}{90} complete\end{turn}}  & Bm & .049 & .075 & .105 & .232 && .014 & .008 & .005 & .001 \\ 
  		& HM1 & .047 & .074 & .101 & .205 && .011 & .006 & .003 & .000 \\ 
  		& HM2 & .051 & .069 & .077 & .120 && .013 & .008 & .004 & .000 \\ 
  		& NW & .040 & .052 & .059 & .108 && .011 & .005 & .002 & .000 \\ 
  		& DL & .051 & .079 & .108 & .212 && .013 & .007 & .004 & .000 \\ 
  		
  		\hline
  		\multirow{5}{*} {\begin{turn}{90} naive\end{turn}}   & Bm & .150 & .290 & .394 & .619 && .025 & .011 & .009 & .009 \\ 
  		& HM1 & .151 & .272 & .366 & .597 && .007 & .001 & .000 & .000 \\ 
  		& HM2 & .164 & .278 & .361 & .577 && .007 & .001 & .000 & .000 \\ 
  		& NW & .111 & .203 & .295 & .569 && .006 & .001 & .000 & .000 \\ 
  		& DL & .170 & .296 & .387 & .607 && .009 & .001 & .000 & .000 \\ 
  		\hline
  		\multirow{5}{*} {\begin{turn}{90} PMM \end{turn}}  & Bm & .031 & .026 & .018 & .005 && .026 & .031 & .031 & .022 \\ 
  		& HM1 & .032 & .027 & .018 & .005 && .023 & .034 & .046 & .070 \\ 
  		& HM2 & .032 & .025 & .014 & .002 && .023 & .035 & .045 & .064 \\ 
  		& NW & .029 & .022 & .013 & .002 && .022 & .029 & .033 & .034 \\ 
  		& DL & .033 & .028 & .019 & .005 && .025 & .037 & .050 & .079 \\ 
  		\hline
  		\multirow{5}{*} {\begin{turn}{90} LR \end{turn}} & Bm & .012 & .005 & .001 & .000 & & .012 & .015 & .015 & .016 \\ 
  		& HM1 & .011 & .005 & .001 & .000& & .012 & .015 & .014 & .013 \\ 
  		& HM2 & .011 & .004 & .000 & .000 && .012 & .015 & .015 & .014 \\ 
  		& NW & .011 & .004 & .000 & .000 && .012 & .014 & .013 & .012 \\ 
  		& DL & .012 & .005 & .001 & .000 && .013 & .015 & .015 & .014 \\ 
  		\hline
  		\multirow{5}{*} {\begin{turn}{90} NORM\end{turn}} & Bm & .007 & .002 & .001 & .000 && .071 & .077 & .115 & .194 \\ 
  		& HM1 & .012 & .004 & .001 & .000 && .062 & .091 & .176 & .356 \\ 
  		& HM2 & .012 & .004 & .001 & .000 && .069 & .103 & .190 & .358 \\ 
  		& NW & .006 & .001 & .000 & .000 && .062 & .077 & .128 & .226 \\ 
  		& DL & .013 & .004 & .001 & .000 && .065 & .095 & .183 & .370 \\ 
  		\hline 
  		
  	\end{tabular}
  \end{table}

\FloatBarrier

\subsection{Performance of CI's for complete data}

Focusing on the top section of Table~\ref{tab:performance2} labeled ``complete", one can see that for each CI method, CP decreases away from the nominal value (0.95) as $\theta$ approaches 1 with its mean absolute error increasing. Among the CI methods, Newcombe's Wald (NW) method has CP noticeably closer to the nominal CP and mean absolute error of CP less than any other methods for all AUC levels. However, since when the true AUC is $0.99$, the CP significantly deviates from the nominal value even for NW method, Wald-type intervals are worth using when AUC is less than equal to $0.95$.
	
Further, as we see in Table~\ref{tab:NCP2}, LNCP's are larger than RNCP's with the imbalance being all the more evident when $\theta$ approaches 1. This is because the symmetric property of Wald-type intervals makes the upper and lower limit over- and underestimated as the standard error decreases as the AUC moves from 0.5 to 1. 

\subsection{Performance of CI's for incomplete data}
Again focusing on Table~\ref{tab:performance2}, the CP's for complete case analysis are, as expected, substantially farther away from the nominal CP (0.95) than those for the complete data.  The CIL's for complete case analysis are on average longer than those for complete data mostly due to loss of information related to the missing data (Figure~\ref{fig:phirho}).

Next, focusing on results where MI techniques were applied to incomplete data, it is easy to see that the MI methods vastly outperform the naive complete case analysis particularly when the value of the AUC approaches 1.  For instance, when AUC is 0.99, the CP for naive analysis range from 0.287 to 0.355 whereas the worst MI procedure (NORM) has CP's ranging from 0.610 to 0.785.  

While the NORM imputation technique certainly outperforms the naive analysis it is inferior to both PMM and LR in terms of CP.  PMM has coverage probabilities that range from 0.896 up to 0.953, much better than NORM and naive analysis, but still underperforming the LR.  LR slightly overestimates coverage probability with all CP's between 0.96 and 0.97 and shows remarkable accuracy in CP across different values of the AUC.  Counterintuitively, the CI's constructed with LR and PMM outperform even the complete data in terms of CP when the AUC is near 1 as seen from Figure~\ref{fig:phirho}. For example, the CP of PMM ranges from 0.896 to 0.953 for $\theta = 0.99$, while that of the complete data does from 0.767 to 0.892.

The performance of NORM is the worst in terms of both CP and CIL. The longest average CIL with the lowest average CP makes NORM the most inefficient approach. Horton et. al. (2003) pointed out that rounding may cause bias in MI and that it might be better to leave the imputed values as they are.  A large bias of NORM is responsible for its high imbalance between LNCP and RNCP.

CI's for the complete data have better CP's for a larger sample; however, the opposite is true for NORM imputation, which has worse CP's for a larger sample as Figure~\ref{fig:CPCIL} shows. When a point estimate is biased, as is the case here, large sample sizes worsen the under-coverage by reducing the standard error estimates.  

\begin{figure}[h]
	\centering
	\includegraphics[width=\textwidth]{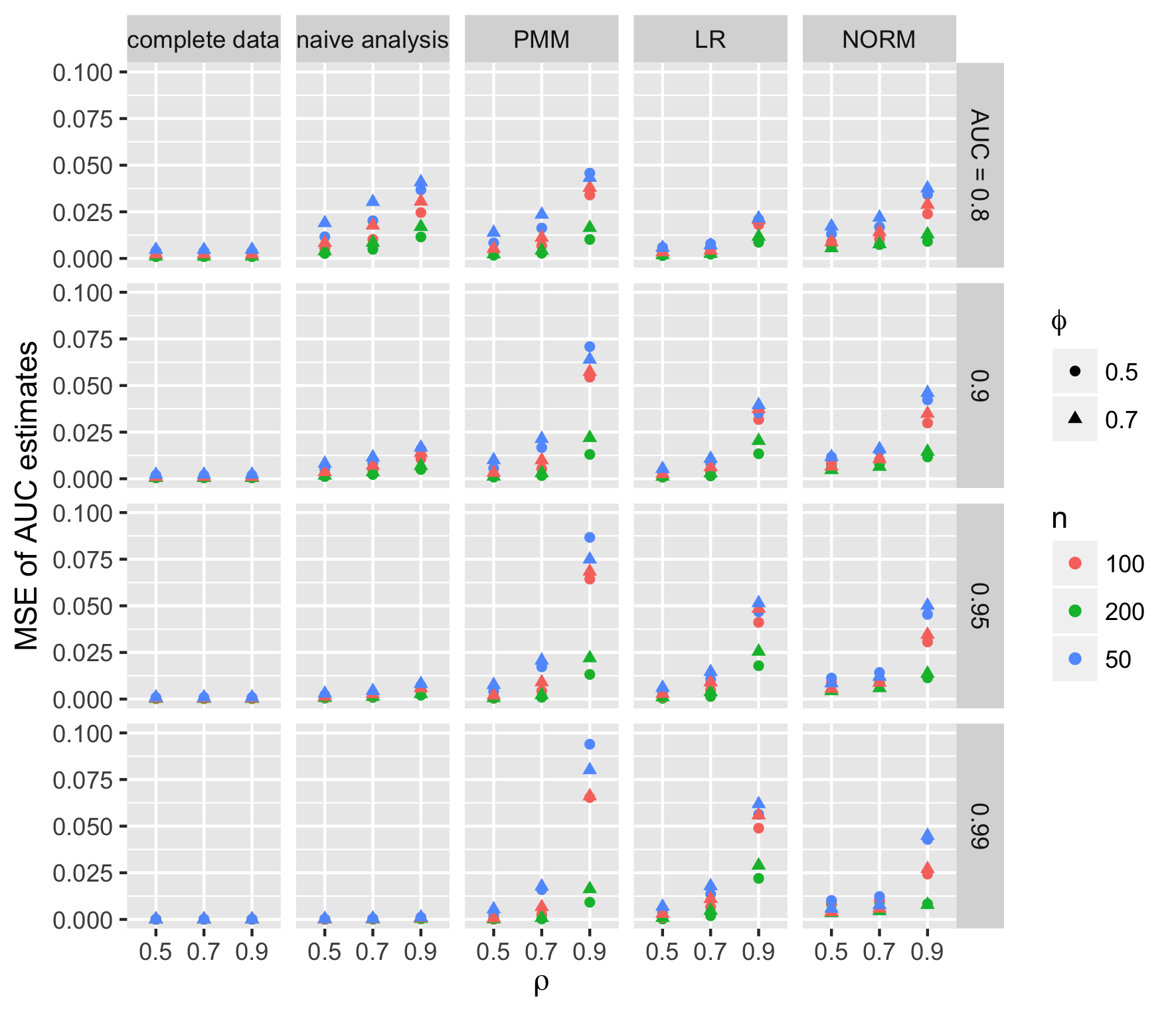}
	\caption{The MSE (mean squared error) of AUC estimates by MI techniques, $\phi$ and $\rho$.}\label{fig:bias_MSE_plot}
\end{figure}

Figure~\ref{fig:bias_MSE_plot} illustrates the mean squared error (MSE) of the point estimates.  We observe that the MSE of naive estimators gets smaller when AUC gets smaller. Because naive estimators overestimate the AUC ($\theta$) under the missingniss mechanism where diseased subjects are more likely to verify their disease status, the bias of naive estimator is bounded by $1-\theta$, which goes to 0 as $\theta$ gets larger. While on the contrary the MI estimators do not necessarily overestimate the AUC, the magnitude of bias can be larger than that of the naive estimator. Also the variance of MI estimators is expected to be larger than that of the naive estimator, under certain settings, the naive estimator outperforms the MI estimators in terms of MSE. However, under moderate levels of AUC ($\theta = 0.8, 0.9$) with moderate level of missingness ($\rho = 0.5, 0.7$), MI, especially LR, reasonably performs well in terms of MSE.

In terms of CIL performance between PMM and LR, PMM tends to yield shorter intervals than LR does, as the population AUC ($\theta$) gets larger. For example, when $\rho=0.70$ and $\theta = 0.99$, the CIL for PMM for range from 0.153 to 0.175 whereas the range of CIL's for LR is from 0.274  to 0.275.  This difference is much less dramatic for smaller values of AUC.  For instance, when $\rho=0.70$ and $\theta = 0.80$, the CIL of PMM ranges from 0.341 to 0.355 whereas with LR the CIL ranges from 0.336 to 0.340.  

Finally, from Figure~\ref{fig:phirho}, we observe that LR gives CP most robust to different settings and closest to the nominal CP. That is, for every values of $phi$ and $\rho$ being considered except $(\phi, \rho) = (0.7, 0.9)$, CP of LR stay close to the nominal CP.	While CIL's of LR are longer than that of PMM, under moderate settings, i.e. $\rho < 0.9, \theta \le 0.95$, its CIL's are comparable to those of PMM. However, when $\rho = 0.5$ (Table~\ref{tab:performance1} in the appendix), NW method with PMM gives the best CP and CIL.

\section{Real data example}
\label{s:example}

To explore the pattern of different CI methods for the AUC when using MI on a real dataset, we examined a group of patients collected by 30 past and present national Alzheimer's Disease Centers (ADCs) and managed by the National Alzheimer's Coordinating Center (NACC) from September 2005 to May 2016.

This longitudinal dataset contains a total of 33,900 subjects with the demographic, clinical (the Uniform Data Set, UDS) and neuropathologic (the Neuropathology Data Set, NP) data collected on subjects, each subject having up to 11 visits. We considered only observations from the initial visits and one of the ADCs. Among these 1014 subjects, $93.4\%$ are not known to have AD and thus their disease status is considered missing, where we defined subjects as diseased if their Alzheimer's disease neuropathologic change (ADNC) score is 1, 2 or 3 (low, intermediate and high ADNC each), and as non-diseased if their ADNC score is 0. Among those who know their AD status of the sample, $83.6\%$ of subjects have AD. 

As a diagnostic test of AD, the Clinical Dementia Rating (CDR) can be used which measures patients' cognitive status with respect to memory, orientation, judgement and problem-solving, community affairs, home and hobbies, and personal care. Each attribute takes on values from 0 to 3 in increments of 0.5 or 1.0, and their sum (CDRSUM), ranges from 0.0 to 18.0. Supplementary variables were also considered for MI of the AD status: age (AGE), sex (SEX), race according to National Institutes of Health (NIH) definitions (RACE), body mass index (BMI), systolic blood pressure (BPSYS), resting heart rate (HRATE), total number of medications reported (NUMMED), years of smoking (SMOKYRS), family history (FAMHIST, 1 if one of the first-degree family has cognitive impairment, 0 otherwise), years of education (EDUCYRS), the total score of Mini-Mental State Examination (MMSE), the total score of Geriatric Depression Score (GDS), and Unified Parkinson's Disease Rating Scale (PDNORMAL). Some of the supplementary variables also have missing values. The descriptive statistics of the variables are summarized in Table~\ref{tab:ADstat}.

\begin{table}[ht]
	\caption{Descriptive statistics of the sample data}
	\centering
	\label{tab:ADstat}
	\begin{tabular}{llccccccc}
		\hline
		variable & type & min & Q1 & median & mean & Q3 & max & \# missing \\ 
		\hline
		CDRSUM & continuous & 0 & 0 & 1.5 & 3.3 & 4.5 & 18 & 0 (0\%)\\
		\multicolumn{2}{l}{ ~CDRSUM (AD) }& 0 & 0.5 & 3.5 & 5.3 & 8.1 & 18 & 0 (0\%)\\
		\multicolumn{2}{l}{ ~CDRSUM (not AD) }& 0  & 0.5  & 2.5 & 3.6 & 3.5 & 18 & 0 (0\%)\\
		AGE & continuous & 29 & 64 & 72 & 71.1 & 79 & 102 & 0 (0\%)\\
		BMI & continuous & 16.5 & 23.1 & 25.9 & 26.5 & 29 & 50.6 & 194 (20.5\%)\\
		BPSYS & continuous & 80 & 122 & 133 & 134.3 & 145 & 213 & 277 (29.3\%) \\
		HRATE & continuous & 39 & 60 & 67 & 67.8 & 74 & 116 & 300 (31.7\%) \\
		NUMMED & continuous & 0 & 3 & 4 & 4.9 & 7 & 22 & 3 (0.3\%)\\
		SMOKYRS & continuous & 0 & 0 & 0 & 8.3 & 10 & 76 & 164 (17.3\%) \\
		EDUCYRS & continuous & 0 & 14 & 16 & 15.8 & 18 & 25 & 1 (0.1\%)\\
		MMSE & continuous & 0 & 21.5 & 27 & 24.3 & 29 & 30 & 139 (14.7\%) \\
		GDS & continuous & 0 & 0 & 1 & 2.2 & 3 & 15 & 107 (11.3\%) \\
		\hline \hline  
		variable & type & \multicolumn{6}{c}{distribution} & \# missing \\ \hline
		AD & binary & \multicolumn{6}{c}{AD: 56 (5.5\%), not AD: 11 (1.1\%)}  & 947 (93.4\%) \\
		SEX & binary & \multicolumn{6}{c}{male: 433 (42.7\%), female: 581 (57.3\%)}  & 0 (0.0\%) \\
		RACE & binary & \multicolumn{6}{c}{ White: 879 (86.7\%), Other 134 (13.2\%)}	& 1 (0.01\%) \\
		FAMHIST & binary & \multicolumn{6}{c}{present: 425 (41.9\%), not present: 511 (50.4\%)}	& 78 (7.7\%) \\
		PDNORMAL & binary & \multicolumn{6}{c}{normal: 374 (36.9\%), not normal: 442 (43.6\%)}	& 198 (19.5\%) \\
		\hline
		\multicolumn{8}{c} {min: minimum, Q1: $1^{\text{st}}$ quartile, Q3: $3^{\text{rd}}$ quartile, max: maximum,}\\ 
		\multicolumn{8}{c} {\# missing: number of missing values}\\
	\end{tabular}
\end{table} 

The naive estimate ($\hat{\theta}_{na}$) of the AUC for the subset is 0.5893, when ignoring the subjects whose AD status is not known. To correct for the verification bias of the estimate, imputations using MICE (PMM and LR) and NORM were performed m = 10 times each.  All the supplementary variables mentioned above were considered in the imputation models. RACE was forced into a binary variable (White versus Other) as multi-categorical variables are not applicable in NORM. Each dataset was analyzed using five variance estimators (Bamber's method (Bm), Hanley-McNeil's method I (HM1), Hanley-McNeil's method II (HM2), Newcombe's Wald method (NW), and DeLong's method (DL)) and by applying Rubin's combining rule, the CI's were constructed.

\begin{figure}[h]
	\centering
	\includegraphics[width=.7\textwidth]{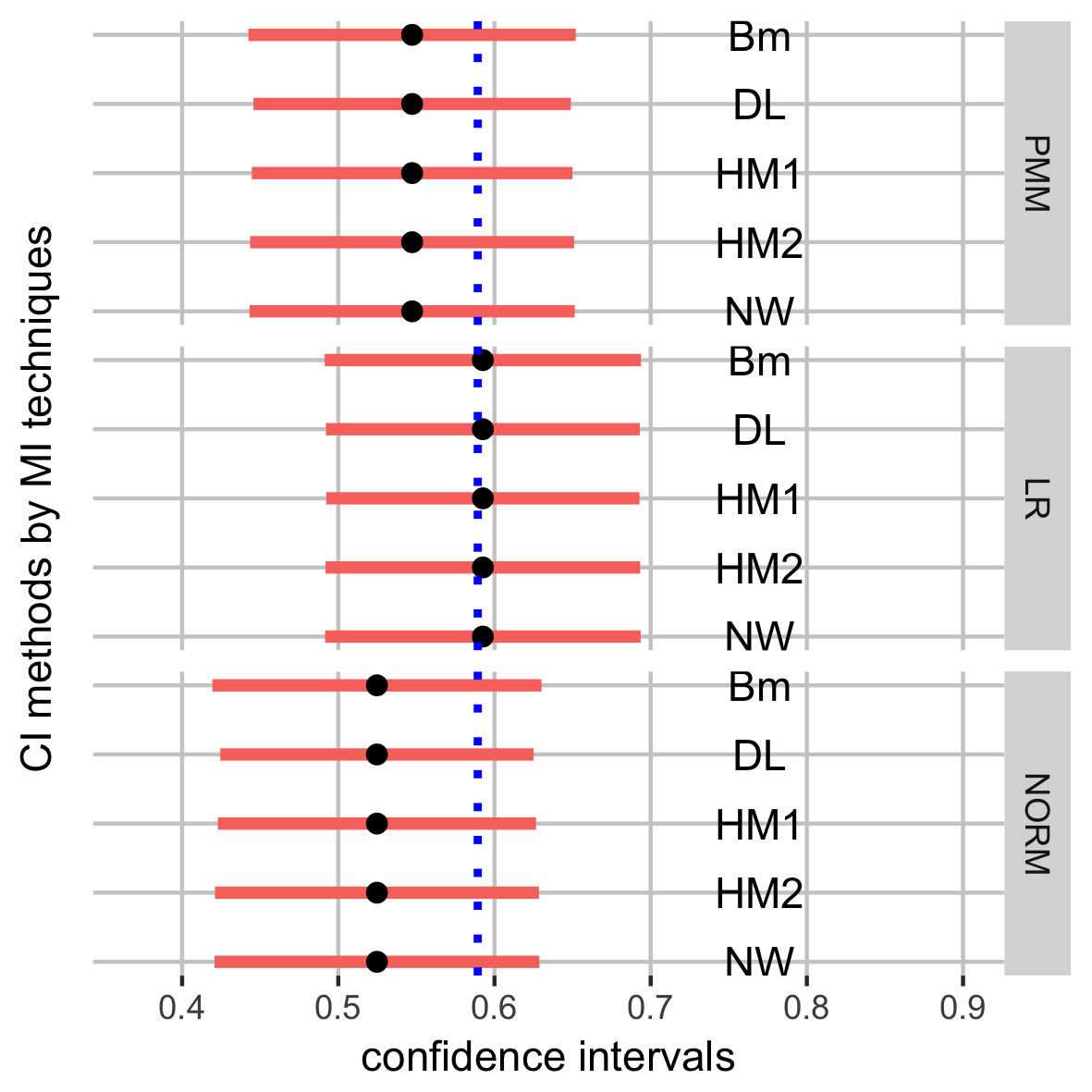}
	\caption{Confidence interval plots of the AUC for CDRSUM as a test of Alzheimer's disease. The blue dotted line is the naive estimate, the black dots in the middle each CI are the bias-corrected estimates by the imputation techniques.}
	\label{fig:AD1}
\end{figure}

The point estimates of the AUC after MI are 0.5473, 0.5926, and 0.5247 for PMM, LR, and NORM respectively. The interval estimates are presented in Figure \ref{fig:AD1}.

Based on the assumption that the probability of AD verification is not dependent on the value of AD status, the point estimate by each MI technique has values lower than or similar to the naive estimate. We can also notice that the CI's, and more specifically their variance estimates, seem to be affected more by the imputation techniques than by the CI methods.

\section{Discussion}
\label{s:discussion}
In this paper, we considered several different methods for constructing Wald-type confidence intervals for the area under the ROC curve in the presence of missingness on disease status indicators with the missingness mechanism assumed to be ignorable.  Several different methods of multiple imputation were considered for handling the missing data, including MICE (PMM and LR) and NORM, as a way to deal with the verification bias problem that arises in some biomedical research where the true disease status is often missing.  We demonstrated through a simulation study that Wald-type CI's, especially NW method, work reasonably well when the true AUC is moderate ($\theta = 0.8, 0.9, 0.95$), that using multiple imputation to handle the missing data greatly outperform a naive complete case analysis confidence interval.  Based on the results we recommend using MI with LR and the choice of CI method is less important. However, when missingness rate is less severe ($< 70\%$), we recommend using NW and MI with PMM.

Such findings are based on some simulation assumptions which may be different from reality to a great degree. However, it is not difficult to analyze data by modifying formulas and to search optimal combination under more realistic settings. For example, we assumed the missing mechanism to be ignorable and the biomarker values to be nearly homoscedastic.  As a solution to the non-ignorable missingness, Harel and Zhou (2006) mentioned that appropriate missingness model can be set up and can be applied in the imputation step, keeping the analysis step and combination step unchanged.  Further, Demirtas and Schafer (2003) and Demirtas (2005), discuss pattern mixture models for imputation when the missingness mechanism is determined to be non-ignorable.  Secondly, the simulation parameters can be set up more comprehensively to find the best method. While our simulation study assumed almost homoscedasticity, there can be biomarkers with significantly different dispersion between groups. In such situation, one can easily design a new simulation study that differentiates the variance of the biomarker.

\section*{Acknowledgements}
The NACC database is funded by NIA/NIH Grant U01 AG016976. NACC data are contributed by the NIAfunded ADCs: P30 AG019610 (PI Eric Reiman, MD), P30 AG013846 (PI Neil Kowall, MD), P50 AG008702 (PI Scott Small, MD), P50 AG025688 (PI Allan Levey, MD, PhD), P50 AG047266 (PI Todd Golde, MD, PhD), P30 AG010133 (PI Andrew Saykin, PsyD), P50 AG005146 (PI Marilyn Albert, PhD), P50 AG005134 (PI Bradley Hyman, MD, PhD), P50 AG016574 (PI Ronald Petersen, MD, PhD), P50 AG005138 (PI Mary Sano, PhD), P30 AG008051 (PI Steven Ferris, PhD), P30 AG013854 (PI M. Marsel Mesulam, MD), P30 AG008017 (PI Jeffrey Kaye, MD), P30 AG010161 (PI David Bennett, MD), P50 AG047366 (PI Victor Henderson, MD, MS), P30 AG010129 (PI Charles DeCarli, MD), P50 AG016573 (PI Frank LaFerla, PhD), P50 AG016570 (PI Marie-Francoise Chesselet, MD, PhD), P50 AG005131 (PI Douglas Galasko, MD), P50 AG023501 (PI Bruce Miller, MD), P30 AG035982 (PI Russell Swerdlow, MD) , P30 AG028383 (PI Linda Van Eldik, PhD), P30 AG010124 (PI John Trojanowski, MD, PhD), P50 AG005133 (PI Oscar Lopez, MD), P50 AG005142 (PI Helena Chui, MD), P30 AG012300 (PI Roger Rosenberg, MD), P50 AG005136 (PI Thomas Montine, MD, PhD), P50 AG033514 (PI Sanjay Asthana, MD, FRCP), P50 AG005681 (PI John Morris, MD), and P50 AG047270 (PI Stephen Strittmatter, MD, PhD).
\vspace*{-8pt}

\appendix
  \section*{Appendix. Proofs}
	\label{A:proofs}
	
	We show that modifications to the Hanley-McNeil's Wald method yields unbiased estimates. First we show how the Hanley-McNeil's variance formula can be derived. Then we prove that the variance estimator is unbiased.
	
	\section{Derivation of Hanley-McNeil's variance formula}	
	Recall that $\theta$, $Q_1$, and $Q_2$ are defined as follows: 
	\begin {equation*}
	\begin {split}
	\theta = & P(Y > X) + \frac{1}{2} P(Y = X), \\
	Q_1 = & P(Y_1, Y_2 > X) + \frac{1}{2} P(Y_1 > Y_2 = X) \\
	& + \frac{1}{2} P(Y_2 > Y_1 = X) + \frac{1}{4} P(Y_1 = Y_2 = X), \\
	Q_2 = & P(Y > X_1, X_2) + \frac{1}{2} P(Y = X_1 > X_2) \\
	& + \frac{1}{2} P(Y = X_2 > X_1) + \frac{1}{4} P(Y = X_1 = X_2). \\
	\end {split}
	\end {equation*}
	
	We also denote that 
	\begin{equation*}
		H_{i,j} = H(Y_i,X_j)=\begin{cases}
			1, & \text{if } Y_i > X_j.\\
			\frac{1}{2}, & \text{if } Y_i = X_j.\\
			0, & \text{otherwise}.
		\end{cases}
	\end{equation*}

	\textit{- Unbiased estimator of the AUC}
	
	Let the AUC estimator be defined as:
	\begin{equation}
		\label{eq:AUChat}
		\hat{\theta} = \frac{1}{n_Xn_Y} \sum_{i,j} H_{i,j}.	
	\end{equation}Then, $E(\hat{\theta})= 1 \cdot P(Y>X) + \frac{1}{2} \cdot P(Y=X) = \theta$.	Thus $\hat{\theta} = \frac{1}{n_Xn_Y} \sum_{i,j} H_{i,j}$ is an unbiased estimator of $\theta$.
	\\
	
	\textit{- Variance of the AUC estimator }
	
	From the equation~(\ref{eq:AUChat}), the variance can be derived as:
	
	\begin {equation*}
	\begin {split}
	V(\hat{\theta}) = &\frac{1}{(n_Xn_Y)^2} \bigg[ \sum_{i,j} V(H_{i,j}) + \sum_{i \neq k} \sum_{j} Cov(H_{i,j}, H_{k,j})\\ 
	& + \sum_{i}\sum_{j \neq h} Cov(H_{i,j}, H_{i,h}) + \sum_{i \neq k}\sum_{j \neq h} Cov(H_{i,j}, H_{k,h})\bigg],\\
	\end {split}
	\end {equation*} 
	where 
	\begin {equation*}
	\begin {split}
	V(H_{i,j}) = & E(H_{i,j}^2) - E(H_{i,j})^2 \\
	= & \{ P(Y>X) + \frac{1}{2^2} P(Y=X) \} - \theta^2 \\
	= & \theta - \frac{1}{4} P(Y=X) - \theta^2,\\
	Cov(H_{i,j}, H_{k,j}) 
	= & E(H_{i,j} H_{k,j}) - E(H_{i,j}) E(H_{k,j}) \\
	= & \{ 1^2 \cdot P(Y_1,Y_2 > X) +  1\cdot\frac{1}{2} \cdot P(Y_2 > Y_1 = X) \\ & +  \frac{1}{2} \cdot 1 \cdot P(Y_1 > Y_2 = X) + \frac{1}{2^2} \cdot P(Y_1=Y_2 = X) \} \\ & - \theta^2 \\
	= & Q_1 - \theta^2,\\ 
	Cov(H_{i,j}, H_{i,h}) 
	= & E(H_{i,j} H_{i,h}) - E(H_{i,j}) E(H_{i,h})\\
	= & Q_2 - \theta^2, \text{ and}\\
	Cov(H_{i,j}, H_{k,h})
	= & E(H_{i,j} H_{k,h}) - E(H_{i,j}) E(H_{k,h}) \\
	= & \theta^2 - \theta^2 \\
	= & 0.\\		
	\end {split}
	\end {equation*}
	
	Then we have:
	\begin {equation}
	\label {eq:var}
	\begin {split}
	V(\hat{\theta}) = & \frac{1}{n_{X}n_{Y}} \bigg[ \theta(1-\theta) - \frac{1}{4} P(Y = X) \\
	& + (n_Y - 1)(Q_1 - \theta^2)+ (n_X - 1)(Q_2 - \theta ^2) \bigg]\\
	\end {split}
	\end {equation}
	
	When the biomarkers are measured continuously enough so that there is no tie, the variance formula (\ref{eq:var}), reduces to what Hanley and McNeil suggested (\ref{eq:var2}):
	
	\begin {equation}
	\label {eq:var2}
	\begin {split}
	V(\hat{\theta}) = & \frac{1}{n_{X}n_{Y}} \bigg[ \theta(1-\theta) + (n_Y - 1)(Q_1 - \theta^2) \\
	& + (n_X - 1)(Q_2 - \theta ^2)\bigg]\\
	\end {split}
	\end {equation}, where $Q_1=P(Y_1,Y_2>X) \text{ and } Q_2=P(Y>X_1,X_2)$.

	\section{Unbiasedness of the modified Hanley-McNeil's variance estimator}
	
	Let the variance estimator be:
	
	\begin {equation}
	\label {eq:varhat}
	\begin {split}
	\widehat{V(\hat{\theta})} = & \frac{1}{(n_{X}-1)(n_{Y}-1)}
	\bigg[\hat{\theta}(1-\hat{\theta}) - \frac{1}{4} p(Y = X) \\
	& + (n_Y - 1)(\hat{Q}_1 - \hat{\theta}^2)
	+ (n_X - 1)(\hat{Q}_2 - \hat{\theta}^2)\bigg],\\
	\end {split}
	\end {equation} where
	
	\begin {equation*}
	\begin {split}
	\hat{Q}_1 = & \frac{1}{n_{X}n_{Y}^{2}}{\sum_{j=1}^{n_X} \bigg[\sum_{i=1}^{n_Y} {H_{i,j}}\bigg]^2} \\
	\hat{Q}_2 = & \frac{1}{n_{X}^2n_{Y}}{\sum_{i=1}^{n_Y} \bigg[\sum_{j=1}^{n_X} {H_{i,j}}\bigg]^2}, \text{ and} \\
	p(Y = X) = & \frac{1}{n_{X}n_{Y}}{\sum_{j=1}^{n_X}\sum_{i=1}^{n_Y}I(y_i = x_j)}\\
	\end {split}
	\end {equation*} 
	
	\textit{- Unbiased estimator of }$Q_1, Q_2, P(Y=X)$
	
	$\hat{Q}_1$, $\hat{Q}_2$, and p(Y=X) are an unbiased estimator of $Q_1$, $Q_2$, and P(Y=X), respectively.\\

	\textit{- Proof }
	
	Let $\{(i,k) \vert i, k =1,2,...,n_Y\}$ be partitioned as:
	\begin {equation*}
	\begin {split}
	A_j = & \{(i,k)|Y_i, Y_k > X_j\},\\
	B_j = & \{(i,k)|Y_i > Y_k = X_j\},\\
	C_j = & \{(i,k)|Y_k > Y_i = X_j\},\\
	D_j = & \{(i,k)|Y_i = Y_k = X_j\}, \text{ and}\\
	E_j = & \{(i,k)|Y_i < X_j \text{ or } Y_k < X_j\}
	\end {split}
	\end {equation*}
	
	Then
	\begin {equation*}
	\begin {split}
	\hat{Q}_1 = & \frac{1}{n_{X}n_{Y}^{2}}{\sum_{j=1}^{n_X} \bigg[\sum_{i=1}^{n_Y} {H_{i,j}}\bigg]^2} 
	= \frac{1}{n_{X}n_{Y}^{2}}{\sum_{j=1}^{n_X} \sum_{i=1}^{n_Y}\sum_{k=1}^{n_Y} H_{i,j}H_{k,j}} \\
	= & \frac{1}{n_{X}n_{Y}^{2}}{\sum_{j=1}^{n_X} \Bigg[
		\sum_{(i,k)\in A_j}^{n_Y} H_{i,j}H_{k,j}}
	+ \sum_{(i,k)\in B_j}^{n_Y} H_{i,j}H_{k,j} \\
	& + \sum_{(i,k)\in C_j}^{n_Y} H_{i,j}H_{k,j}
	+ \sum_{(i,k)\in D_j}^{n_Y} H_{i,j}H_{k,j}
	+ \sum_{(i,k)\in E_j}^{n_Y} H_{i,j}H_{k,j}
	\Bigg]\\
	= & \frac{1}{n_{X}n_{Y}^{2}}{\sum_{j=1}^{n_X} \Bigg[
		\sum_{(i,k)\in A_j}^{n_Y} 1 \cdot 1}
	+ \sum_{(i,k)\in B_j}^{n_Y} 1 \cdot \frac{1}{2}
	+ \sum_{(i,k)\in C_j}^{n_Y} \frac{1}{2} \cdot 1 \\
	& + \sum_{(i,k)\in D_j}^{n_Y} \frac{1}{2} \cdot \frac{1}{2}
	+ \sum_{(i,k)\in E_j}^{n_Y} 0
	\Bigg]\\
	\end {split}
	\end {equation*}
	
	\begin {equation*}
	\begin {split}
	E[\hat{Q}_1]
	= & \frac{1}{n_{X}n_{Y}^{2}}E\Bigg[\sum_{j=1}^{n_X} 
	\sum_{(i,k)\in A_j}^{n_Y} 1
	+ \sum_{j=1}^{n_X}\sum_{(i,k)\in B_j}^{n_Y} \frac{1}{2}
	+ \sum_{j=1}^{n_X}\sum_{(i,k)\in C_j}^{n_Y} \frac{1}{2} \\
	& + \sum_{j=1}^{n_X}\sum_{(i,k)\in D_j}^{n_Y} \frac{1}{4}
	\Bigg]\\
	= & \frac{1}{n_{X}n_{Y}^{2}}\{n_X{n_Y}^2 P(Y_1,Y_2 > X)
	+ \frac{1}{2} n_X{n_Y}^2 P(Y_1 > Y_2 = X) \\
	& + \frac{1}{2} n_X{n_Y}^2 P(Y_2 > Y_1 = X)
	+ \frac{1}{4} n_X{n_Y}^2 P(Y_1 = Y_2 = X)\}\\
	= & P(Y_1,Y_2 > X) + P(Y_1 > Y_2 = X) \\
	& + P(Y_2 > Y_1 = X) + P(Y_1 = Y_2 = X)\}\\
	= & Q_1\\
	\end {split}
	\end {equation*}
	
	$\hat{Q}_2$ can be shown to be unbiased in a similar way. Also $E[p(Y=X)]=E[\frac{1}{n_{X}n_{Y}}{\sum_{j=1}^{n_X}\sum_{i=1}^{n_Y}I(y_i = x_j)}] = \frac{1}{n_{X}n_{Y}}{n_{X}n_{Y}P(Y=X)}=P(Y=X)$.\\

	\textit{- Unbiased estimator of } $V(\hat{\theta})$
	
	$\widehat{V(\hat{\theta})}$ defined in (\ref{eq:varhat}) is an unbiased estimator of $V(\hat{\theta})$.\\
	
	\textit{- Proof}
	\begin {equation*}
	\begin {split}
	E[\widehat{V(\hat{\theta})}] = & 
	\frac{1}{(n_{X}-1)(n_{Y}-1)} E\Bigg[ \hat{\theta}(1-\hat{\theta}) - \frac{1}{4} p(Y = X) \\ 
	& ~~+ (n_Y - 1)(\hat{Q}_1 - \hat{\theta}^2)+ (n_X - 1)(\hat{Q}_2 - \hat{\theta}^2)\Bigg] \\
	= & \frac{1}{(n_{X}-1)(n_{Y}-1)} \Bigg[ E[\hat{\theta}(1-\hat{\theta})] - E[\frac{1}{4} p(Y = X)]\\
	& ~~+ (n_Y - 1)E[\hat{Q}_1 - \hat{\theta}^2]+ (n_X - 1)E[\hat{Q}_2 - \hat{\theta}^2] \Bigg]\\
	= & \frac{1}{(n_{X}-1)(n_{Y}-1)} \Bigg[\theta - \theta^2 - \frac{1}{4} P(Y = X) \\ & ~~+ (n_Y - 1)(Q_1 - \theta^2) + (n_X - 1)(Q_1 - \theta^2) \\ & ~~- (n_X + n_Y - 1)V(\hat{\theta}) \Bigg] \\
	= & \frac{n_Xn_YV(\hat{\theta}) - (n_X + n_Y - 1)V(\hat{\theta})}{(n_{X} - 1)(n_{Y} - 1)} \\
	= & \frac{(n_X - 1)(n_Y-1)V(\hat{\theta})} {(n_{X} - 1)(n_{Y} - 1)} \\
	= & V(\hat{\theta}) \\
	\end {split}
	\end {equation*}
	
\section*{Appendix. Additional Tables}

\begin{table}[ht]
	\caption{Performance of CI's for each imputation and CI method when $\rho = 50\%$} \label{tab:performance1}
	\centering
	\begin{tabular}{ll cccc cccc cccc}
		\hline
		MI & CI &
		\multicolumn{4}{c}{CP} &
		\multicolumn{4}{c}{MAE (CP)} &
		\multicolumn{4}{c}{CIL} \\
		\multicolumn{2}{r}{}  & $\theta$ = 0.8 & 0.9 & 0.95 & 0.99  & 0.8 & 0.9 & 0.95 & 0.99 & 0.8 & 0.9 & 0.95 & 0.99 \\ 
		\hline
		\hline
		\multirow{5}{*} {\begin{turn}{90} complete\end{turn}} & Bm & .934 & .917 & .889 & .763 & .014 & .033 & .061 & .184 & .187 & .126 & .079 & .022 \\ 
		& HM1 & .938 & .921 & .896 & .791 & .010 & .030 & .054 & .157 & .192 & .130 & .082 & .023 \\ 
		& HM2 & .932 & .925 & .918 & .877 & .017 & .026 & .035 & .076 & .185 & .128 & .084 & .028 \\ 
		& NW & .945 & .943 & .940 & .888 & .005 & .010 & .017 & .070 & .195 & .137 & .091 & .030 \\ 
		& DL & .934 & .915 & .889 & .783 & .015 & .035 & .062 & .165 & .187 & .127 & .080 & .023 \\ 
		\hline
		\multirow{5}{*} {\begin{turn}{90} naive\end{turn}} & Bm & .846 & .757 & .681 & .479 & .287 & .385 & .469 & .659 & .279 & .165 & .091 & .020 \\ 
		& HM1 & .886 & .796 & .723 & .516 & .177 & .299 & .403 & .629 & .316 & .185 & .103 & .024 \\ 
		& HM2 & .866 & .785 & .734 & .559 & .188 & .304 & .395 & .608 & .289 & .169 & .097 & .025 \\ 
		& NW & .909 & .856 & .805 & .584 & .149 & .251 & .349 & .597 & .320 & .198 & .117 & .032 \\ 
		& DL & .865 & .771 & .698 & .499 & .239 & .347 & .437 & .642 & .300 & .173 & .094 & .021 \\ 
		\hline
		\multirow{5}{*} {\begin{turn}{90} PMM\end{turn}} & Bm & .935 & .933 & .940 & .955 & .064 & .100 & .119 & .127 & .271 & .203 & .150 & .082 \\ 
		& HM1 & .938 & .933 & .939 & .953 & .062 & .105 & .130 & .153 & .271 & .198 & .143 & .074 \\ 
		& HM2 & .937 & .937 & .948 & .976 & .061 & .102 & .125 & .150 & .268 & .198 & .145 & .077 \\ 
		& NW & .943 & .946 & .958 & .981 & .058 & .096 & .115 & .133 & .276 & .207 & .153 & .082 \\ 
		& DL & .935 & .930 & .937 & .950 & .064 & .109 & .134 & .158 & .268 & .196 & .142 & .073 \\ 
		\hline
		\multirow{5}{*} {\begin{turn}{90} LR \end{turn}} & Bm & .960 & .973 & .982 & .991 & .036 & .046 & .054 & .059 & .261 & .214 & .182 & .152 \\ 
		& HM1 & .962 & .975 & .983 & .993 & .036 & .047 & .054 & .060 & .263 & .216 & .183 & .153 \\ 
		& HM2 & .961 & .976 & .984 & .992 & .036 & .047 & .054 & .059 & .261 & .215 & .184 & .153 \\ 
		& NW & .964 & .979 & .987 & .995 & .037 & .049 & .055 & .060 & .265 & .219 & .187 & .155 \\ 
		& DL & .960 & .973 & .982 & .993 & .036 & .046 & .054 & .060 & .260 & .214 & .182 & .152 \\ 
		\hline
		\multirow{5}{*} {\begin{turn}{90} NORM\end{turn}} & Bm & .900 & .886 & .845 & .792 & .118 & .145 & .193 & .255 & .317 & .270 & .235 & .188 \\ 
		& HM1 & .910 & .873 & .782 & .614 & .114 & .158 & .235 & .373 & .319 & .254 & .208 & .158 \\ 
		& HM2 & .899 & .853 & .764 & .617 & .121 & .170 & .250 & .379 & .307 & .246 & .206 & .159 \\ 
		& NW & .913 & .890 & .834 & .750 & .110 & .146 & .201 & .278 & .325 & .269 & .228 & .178 \\ 
		& DL & .907 & .868 & .775 & .600 & .117 & .161 & .241 & .386 & .314 & .250 & .205 & .155 \\ 
		\hline
		\hline
	\end{tabular}
\end{table}

\begin{table}[ht]
	\caption{Performance of CI's for each imputation and CI method when $\rho = 90\%$} \label{tab:performance3}
	\centering
	\begin{tabular}{llcccc cccc cccc}
		\hline
		MI & CI &
		\multicolumn{4}{c}{CP} &
		\multicolumn{4}{c}{MAE (CP)} &
		\multicolumn{4}{c}{CIL} \\
		\multicolumn{2}{r}{}  & $\theta$ = 0.8 & 0.9 & 0.95 & 0.99  & 0.8 & 0.9 & 0.95 & 0.99 & 0.8 & 0.9 & 0.95 & 0.99 \\ 
		\hline
		\multirow{5}{*} {\begin{turn}{90} complete\end{turn}} & Bm & .936 & .917 & .888 & .766 & .014 & .033 & .062 & .184 & .187 & .126 & .080 & .022 \\ 
		& HM1 & .939 & .921 & .896 & .794 & .011 & .029 & .054 & .156 & .192 & .130 & .082 & .024 \\ 
		& HM2 & .933 & .925 & .919 & .879 & .017 & .026 & .036 & .076 & .185 & .128 & .084 & .028 \\ 
		& NW & .946 & .943 & .940 & .890 & .005 & .010 & .017 & .071 & .195 & .137 & .091 & .030 \\ 
		& DL & .934 & .915 & .888 & .786 & .016 & .035 & .062 & .164 & .187 & .127 & .080 & .023 \\ 
		\hline
		\multirow{5}{*} {\begin{turn}{90} naive\end{turn}} & Bm & .449 & .364 & .274 & .107 & .501 & .586 & .676 & .843 & .344 & .178 & .088 & .018 \\ 
		& HM1 & .626 & .483 & .351 & .121 & .324 & .467 & .599 & .829 & .490 & .296 & .169 & .041 \\ 
		& HM2 & .628 & .487 & .357 & .121 & .322 & .463 & .593 & .829 & .485 & .293 & .167 & .041 \\ 
		& NW & .647 & .500 & .360 & .121 & .303 & .450 & .590 & .829 & .506 & .311 & .180 & .045 \\ 
		& DL & .518 & .415 & .311 & .113 & .432 & .535 & .639 & .837 & .372 & .199 & .104 & .022 \\ 
		\hline
		\multirow{5}{*} {\begin{turn}{90} PMM \end{turn}} & Bm & .799 & .693 & .636 & .617 & .151 & .257 & .314 & .333 & .506 & .479 & .454 & .410 \\ 
		& HM1 & .799 & .684 & .615 & .568 & .151 & .266 & .335 & .382 & .498 & .470 & .444 & .395 \\ 
		& HM2 & .804 & .688 & .619 & .574 & .146 & .262 & .331 & .376 & .503 & .473 & .446 & .396 \\ 
		& NW & .806 & .698 & .636 & .605 & .144 & .252 & .314 & .345 & .509 & .481 & .455 & .407 \\ 
		& DL & .795 & .678 & .609 & .560 & .155 & .272 & .341 & .389 & .495 & .466 & .440 & .392 \\ 
		\hline
		\multirow{5}{*} {\begin{turn}{90} LR\end{turn}} & Bm & .883 & .871 & .865 & .862 & .072 & .087 & .095 & .101 & .574 & .560 & .551 & .539 \\ 
		& HM1 & .884 & .872 & .866 & .863 & .072 & .087 & .094 & .099 & .576 & .561 & .552 & .540 \\ 
		& HM2 & .884 & .872 & .866 & .862 & .072 & .087 & .094 & .100 & .575 & .561 & .552 & .539 \\ 
		& NW & .884 & .872 & .866 & .863 & .072 & .087 & .094 & .099 & .576 & .561 & .552 & .540 \\ 
		& DL & .883 & .872 & .865 & .862 & .072 & .087 & .095 & .100 & .575 & .560 & .551 & .539 \\ 
		\hline
		\multirow{5}{*} {\begin{turn}{90} NORM \end{turn}} & Bm & .694 & .642 & .594 & .559 & .256 & .308 & .356 & .391 & .323 & .289 & .259 & .207 \\ 
		& HM1 & .693 & .635 & .574 & .502 & .257 & .315 & .376 & .448 & .320 & .282 & .251 & .196 \\ 
		& HM2 & .688 & .624 & .559 & .487 & .262 & .326 & .391 & .463 & .315 & .278 & .246 & .193 \\ 
		& NW & .701 & .649 & .599 & .559 & .249 & .301 & .351 & .391 & .327 & .292 & .261 & .207 \\ 
		& DL & .688 & .629 & .566 & .489 & .262 & .321 & .384 & .461 & .315 & .278 & .247 & .193 \\ 
		\hline
	\end{tabular}
\end{table}

\begin{table}[ht]
	\caption{Non-coverage probabilities of CI's for each imputation and CI method when $\rho = 50\%$} \label{tab:NCP1}
	\centering
	\begin{tabular}{llcccc p{0.01cm} cccc}
		\hline
		MI & CI &
		\multicolumn{4}{c}{LNCP} &&
		\multicolumn{4}{c}{RNCP} \\
		\multicolumn{2}{r}{}  & $\theta$ = 0.8 & 0.9 & 0.95 & 0.99  && 0.8 & 0.9 & 0.95 & 0.99 \\ 
		\hline
		\multirow{5}{*} {\begin{turn}{90} complete\end{turn}}  & Bm & .052 & .075 & .105 & .236 && .014 & .008 & .005 & .001 \\ 
		& HM1 & .050 & .073 & .100 & .209 && .011 & .007 & .004 & .001 \\ 
		& HM2 & .054 & .066 & .077 & .122 && .014 & .008 & .005 & .000 \\ 
		& NW & .044 & .051 & .058 & .112 && .012 & .006 & .002 & .000 \\ 
		& DL & .053 & .077 & .107 & .217 && .013 & .007 & .004 & .001 \\ 
		\hline
		\multirow{5}{*} {\begin{turn}{90} naive\end{turn}}   & Bm & .094 & .196 & .284 & .496 && .026 & .011 & .007 & .007 \\ 
		& HM1 & .099 & .193 & .265 & .471 && .011 & .002 & .000 & .000 \\ 
		& HM2 & .115 & .203 & .254 & .428 && .015 & .004 & .001 & .000 \\ 
		& NW & .075 & .134 & .184 & .402 && .012 & .002 & .000 & .000 \\ 
		& DL & .110 & .212 & .286 & .486 && .013 & .003 & .001 & .000 \\ 
		\hline
		\multirow{5}{*} {\begin{turn}{90} PMM \end{turn}}  & Bm & .044 & .050 & .048 & .041 && .018 & .013 & .009 & .002 \\ 
		& HM1 & .044 & .051 & .048 & .040 && .016 & .013 & .010 & .005 \\ 
		& HM2 & .044 & .046 & .038 & .018 && .017 & .014 & .012 & .004 \\ 
		& NW & .039 & .040 & .032 & .015 && .015 & .011 & .007 & .001 \\ 
		& DL & .046 & .053 & .049 & .042 && .017 & .014 & .011 & .006 \\ 
		\hline
		\multirow{5}{*} {\begin{turn}{90} LR \end{turn}}  & Bm & .027 & .015 & .007 & .000 && .011 & .009 & .009 & .006 \\ 
		& HM1 & .026 & .014 & .006 & .000 && .010 & .009 & .008 & .004 \\ 
		& HM2 & .025 & .011 & .004 & .000 && .011 & .010 & .010 & .005 \\ 
		& NW & .024 & .011 & .004 & .000 && .011 & .008 & .007 & .003 \\ 
		& DL & .027 & .014 & .007 & .000 && .011 & .009 & .009 & .005 \\ 
		\hline
		\multirow{5}{*} {\begin{turn}{90} NORM\end{turn}} & Bm & .004 & .001 & .000 & .000 && .090 & .108 & .151 & .205 \\ 
		& HM1 & .007 & .002 & .000 & .000 && .081 & .122 & .214 & .383 \\ 
		& HM2 & .007 & .002 & .000 & .000 && .092 & .142 & .233 & .381 \\ 
		& NW & .004 & .001 & .000 & .000 && .081 & .106 & .163 & .248 \\ 
		& DL & .007 & .003 & .000 & .000 && .084 & .126 & .222 & .398 \\ 
		\hline 
	\end{tabular}
\end{table}

\begin{table}[ht]
	\caption{Non-coverage probabilities of CI's for each imputation and CI method when $\rho = 90\%$} \label{tab:NCP3}
	\centering
	\begin{tabular}{llcccc p{0.01cm} cccc}
		\hline
		MI & CI &
		\multicolumn{4}{c}{LNCP} &&
		\multicolumn{4}{c}{RNCP} \\
		\multicolumn{2}{r}{}  & $\theta$ = 0.8 & 0.9 & 0.95 & 0.99  && 0.8 & 0.9 & 0.95 & 0.99 \\ 
		\hline
		\multirow{5}{*} {\begin{turn}{90} complete\end{turn}}  & Bm & .051 & .074 & .106 & .233  && .014 & .008 & .005 & .001 \\ 
		& HM1 & .049 & .073 & .101 & .205  && .012 & .007 & .003 & .000 \\ 
		& HM2 & .053 & .067 & .077 & .121 & & .014 & .008 & .005 & .000 \\ 
		& NW & .043 & .051 & .058 & .110  && .011 & .006 & .002 & .000 \\ 
		& DL & .053 & .077 & .108 & .214  && .013 & .007 & .004 & .001 \\ 
		\hline
		\multirow{5}{*} {\begin{turn}{90} naive\end{turn}}  & Bm & .226 & .323 & .421 & .608 & & .006 & .007 & .012 & .004 \\ 
		& HM1 & .183 & .289 & .399 & .608  && .002 & .000 & .000 & .000 \\ 
		& HM2 & .183 & .285 & .392 & .608  && .001 & .000 & .000 & .000 \\ 
		& NW & .164 & .272 & .390 & .608  && .001 & .000 & .000 & .000 \\ 
		& DL & .202 & .302 & .407 & .608  && .005 & .001 & .000 & .000 \\ 
		\hline
		\multirow{5}{*} {\begin{turn}{90} PMM \end{turn}}  & Bm & .004 & .001 & .000 & .000 & & .103 & .211 & .268 & .286 \\ 
		& HM1 & .005 & .001 & .000 & .000  && .105 & .223 & .291 & .337 \\ 
		& HM2 & .004 & .001 & .000 & .000  && .100 & .219 & .287 & .331 \\ 
		& NW & .004 & .001 & .000 & .000  && .099 & .209 & .270 & .300 \\ 
		& DL & .005 & .001 & .000 & .000  && .109 & .229 & .297 & .345 \\ 
		\hline
		\multirow{5}{*} {\begin{turn}{90} LR \end{turn}}  & Bm & .001 & .000 & .000 & .000 & & .025 & .037 & .041 & .044 \\ 
		& HM1 & .001 & .000 & .000 & .000  && .024 & .036 & .040 & .042 \\ 
		& HM2 & .001 & .000 & .000 & .000  && .024 & .036 & .041 & .043 \\ 
		& NW & .001 & .000 & .000 & .000  && .024 & .036 & .041 & .043 \\ 
		& DL & .001 & .000 & .000 & .000  && .025 & .036 & .041 & .043 \\ 
		\hline
		\multirow{5}{*} {\begin{turn}{90} NORM\end{turn}} & Bm & .036 & .012 & .005 & .002  && .178 & .250 & .304 & .345 \\ 
		& HM1 & .040 & .014 & .005 & .002  && .177 & .257 & .325 & .404 \\ 
		& HM2 & .042 & .014 & .005 & .001  && .180 & .267 & .340 & .419 \\ 
		& NW & .034 & .010 & .003 & .001  && .174 & .247 & .303 & .347 \\ 
		& DL & .041 & .015 & .006 & .002  && .181 & .261 & .332 & .417 \\ 
		\hline

	\end{tabular}
\end{table}

\vspace*{-8pt}

%


\label{lastpage}

\end{document}